\documentclass[prl,aps,nofootinbib,notitlepage,twocolumn,superscriptaddress]{revtex4-1}

\usepackage{amsthm}
\usepackage{amsmath}
\usepackage{amssymb}
\usepackage{amsfonts}
\usepackage{graphicx}

\usepackage{txfonts}

\usepackage{ulem}
\usepackage{verbatim}

\usepackage[colorlinks=true,linkcolor=blue,citecolor=blue,urlcolor=blue]{hyperref}

\begin{document}

\title{Metrological Nonlinear Squeezing Parameter}
\author{Manuel Gessner}
\affiliation{QSTAR, CNR-INO and LENS, Largo Enrico Fermi 2, 50125 Firenze, Italy}
\affiliation{D\'{e}partement de Physique, \'{E}cole Normale Sup\'{e}rieure, PSL Universit\'{e}, CNRS,
24 Rue Lhomond, 75005 Paris, France}
\author{Augusto Smerzi}
\affiliation{QSTAR, CNR-INO and LENS, Largo Enrico Fermi 2, 50125 Firenze, Italy}
\author{Luca Pezz\`e}
\affiliation{QSTAR, CNR-INO and LENS, Largo Enrico Fermi 2, 50125 Firenze, Italy}
\date{\today}

\begin{abstract}
The well known metrological linear squeezing parameters (such as quadrature or spin squeezing)
efficiently quantify the sensitivity of Gaussian states. 
Yet, these parameters are insufficient to characterize the much wider class of highly sensitive non-Gaussian states. 
Here, we introduce a class of metrological nonlinear squeezing parameters obtained by analytical optimization of measurement observables 
among a given set of accessible (possibly nonlinear) operators. This allows for the metrological characterization of non-Gaussian quantum states of discrete and continuous variables. Our results lead to optimized and experimentally-feasible recipes for high-precision moment-based estimation of a phase parameter and can be used to systematically construct multipartite entanglement and non-classicality witnesses for complex quantum states.
\end{abstract}

\maketitle

\textit{Introduction.}---A central quest in quantum metrology is to relate the reduced variance of an observable to the possible enhancement of sensitivity in parameter estimation~\cite{Caves, Wineland,RMP,TothJPA}. 
For instance, quadrature squeezing can enhance the sensitivity of homodyne interferometers beyond the shot-noise limit \cite{Caves}, 
as experimentally demonstrated with squeezed vacuum states of light~\cite{GrangierPRL1987, Kimble} and atoms~\cite{KrusePRL2016}, and envisaged for third-generation gravitational wave detectors~\cite{SchnabelNATCOMM2010, AasiNATPHOT2013}. 
Moreover, multi-mode squeezing can reveal mode entanglement~\cite{Adesso, Quantum2017, GrossNATURE,vanLoock,Furusawa,Treps} and Einstein-Podolski-Rosen correlations~\cite{ReidRMP2009,EPRoptical,EPRatomic, FadelSCIENCE2018}.
Squeezing of a collective spin~\cite{Wineland} currently represents the leading strategy to obtain quantum-enhanced sensitivities in Ramsey interferometers~\cite{RMP}, with direct applications to atomic clocks \cite{LudlowRMP2015}, magnetometers \cite{Mitchell}, and matter-wave interferometers \cite{Pritchard}. 
Spin squeezing is also a witness of metrologically-useful multiparticle-entanglement~\cite{Sorensen, SMPRL01,Toth, HyllusPRA2012b} and 
Bell correlations~\cite{Tura,Schmied,EngelsenPRL2017}. Squeezing of linear observables of discrete \cite{KitagawaUeda, Ma, Toth, Sorensen, SMPRL01, HyllusPRA2012b, Sinatra, Bohnet, LocalSqueezing, Riedel, Hosten, LerouxPRL2010} or continuous variables 
\cite{BraunsteinVanLoock, Ferraro, Wang, Weedbrook,Lang}, 
e.g., collective spins or quadratures, 
has proven to be a successful concept to characterize the class of 
Gaussian quantum states with phase-estimation sensitivities beyond the classical limit \cite{RMP}, 
and is hereinafter indicated as metrological linear squeezing. 

Yet, some highly sensitive continuous-variable states are non-Gaussian~\cite{Zurek,Ourjoumtsev,DelegliseNATURE2008,Wolf} and Gaussian spin states form a small and non-optimal class of useful states for quantum metrology~\cite{PS09,Varenna,GiovannettiNATPHOT2011}.
Non-Gaussian states further hold the promise of opening up classically intractable pathways for quantum information processing \cite{Bartlett,Eisert,Giedke}. 
These perspectives have led to a growing interest in the generation of non-Gaussian quantum states in both discrete- and continous-variable systems \cite{Andersen} using nonlinear processes \cite{Strobel,Sciarrino}, photon-addition or -subtraction \cite{Grangier,Bellini,Walschaers} or measurement techniques~\cite{Ourjoumtsev,DelegliseNATURE2008,HaasSCIENCE2014}. 
More refined tools are required to characterize highly sensitive non-Gaussian states, as the 
linear squeezing coefficient becomes too coarse to capture non-Gaussian features~\cite{Strobel}. 
It would be highly desirable to reveal the metrological sensitivity of non-Gaussian states using only the mean value and variance of some accessible nonlinear observables, beyond the limitations of linear squeezing. However, this possibility has been demonstrated only for specific cases---e.g., using squared spin operators for twin-Fock states~\cite{KimPRA1998,TF}, or the parity operator for GHZ states~\cite{BollingerPRA1996, LeibfriedNATURE2005, MonzPRL2011}---and it is not known how to systematically identify optimal nonlinear observables for arbitrary states. While in principle, we may theoretically determine an optimal projective measurement that 
will fully reveal the metrological potential of any quantum state \cite{BraunsteinPRL1994}, 
such a measurement is experimentally unfeasible in most cases.

In this manuscript, we extend the concept of metrological linear squeezing to arbitrary (nonlinear) observables and provide a systematic way to optimize it.
Specifically, we analytically identify the optimal measurement observable out of any given family of accessible operators
for arbitrary quantum states. Measurement of this observable will lead to the highest achievable metrological sensitivity in a quantum phase estimation experiment within this family of accessible operators.
If this family includes only linear observables, we recover the well known linear squeezing parameters.
If also nonlinear operators are accessible, we obtain metrological nonlinear squeezing parameters 
that are suitable to characterize the sensitivity of a wider class of
non-Gaussian quantum states, as we illustrate with a series of examples. 
When all possible measurement operators are accessible, the ensemble of states detected by the nonlinear squeezing parameters equals the 
full ensemble of metrologically-useful quantum states detected by the quantum Fisher information. 
Our results provide scalable tools for the development of feasible quantum phase estimation strategies beyond Gaussian states and the identification of multiparticle entanglement in increasingly complex many-body quantum systems.

\textit{Metrological nonlinear squeezing parameter}.---One possibility to estimate an unknown parameter $\theta$ encoded in a quantum state $\hat{\rho}(\theta)$ is given by the method of moments~(see \cite{Varenna} for a review).
Within this approach $\theta$ is estimated as the parameter $\theta_{\rm est}$ 
that yields equality of the sample mean $\bar{x}_\mu = \sum_{i=1}^\mu x_i/\mu$ of a sequence of independent measurements
and the expectation value $\langle \hat{X}\rangle_{\hat{\rho}(\theta)}=\mathrm{Tr}\{\hat{X}\hat{\rho}(\theta)\}$ 
that is determined in a calibration experiment beforehand, i.e., $\bar{x}_\mu = \langle \hat{X}\rangle_{\hat{\rho}(\theta_{\rm est})}$. In the central limit, for $\mu$ sufficiently large, the random variable $\bar{x}_\mu$ is normally distributed 
around $\langle \hat{X}\rangle_{\hat{\rho}(\theta)}$
with variance $(\Delta\hat{X})_{\hat{\rho}(\theta)}^2/\mu$. 
Then, the phase uncertainty is given by 
$(\Delta\theta_{\mathrm{est}})^2= \chi^2[\hat{\rho}(\theta),\hat{X}]/\mu$,
where 
$\chi^2[\hat{\rho}(\theta),\hat{X}] = (\Delta\hat{X})^2_{\hat{\rho}(\theta)}(\frac{d\langle \hat{X} \rangle_{\hat{\rho}(\theta)}}{d\theta})^{-2}$
is the squeezing parameter of $\hat{\rho}$ associated with the measurement of the observable $\hat{X}$. It can be obtained by error propagation and quantifies the squeezing of the measurement variance $(\Delta\hat{X})^2_{\hat{\rho}(\theta)}$ with respect to the variation of the expectation value $\langle \hat{X}\rangle_{\hat{\rho}(\theta)}$ with $\theta$. 
The squeezing parameter fulfills the chain of inequalities 
$\chi^{-2}[\hat{\rho}(\theta),\hat{X}]\leq F[\hat{\rho}(\theta),\hat{X}]\leq F_Q[\hat{\rho},\hat{H}]$ that is saturable by an optimal measurement observable $\hat{X}$ \cite{Varenna,BraunsteinPRL1994,Frowis2015,Kholevo}.
Here, $F[\hat{\rho}(\theta),\hat{X}]=\sum_xp(x|\theta)(\frac{d}{d\theta}\log p(x|\theta))^2$ is the Fisher information, where 
$p(x|\theta)=\mathrm{Tr}\{\hat{\Pi}_x\hat{\rho}(\theta)\}$ describes the full counting statistics for the observable $\hat{X}$ with spectral decomposition $\hat{X}=\sum_x x\hat{\Pi}_x$. Finally, 
the quantum Fisher information $F_Q[\hat{\rho},\hat{H}]=\max_{\hat{X}}F[\hat{\rho}(\theta),\hat{X}]$ defines the state's full metrological potential by providing the quantum Cram\'{e}r-Rao sensitivity limit $(\Delta\theta_{\mathrm{est}})^2\geq (\mu F_Q[\hat{\rho},\hat{H}])^{-1}$ for arbitrary unbiased estimation strategies \cite{note1,BraunsteinPRL1994,HelstromBOOK}. For unitary evolutions $\hat{\rho}(\theta)=e^{-i\hat{H}\theta}\hat{\rho}e^{i\hat{H}\theta}$ generated by $\hat{H}$, the parameter \cite{note2}
\begin{align}\label{eq:mmoments2}
\chi^2[\hat{\rho},\hat{H},\hat{X}]=\frac{(\Delta\hat{X})^2_{\hat{\rho}}}{\vert \langle [\hat{X},\hat{H}] \rangle_{\hat{\rho}} \vert^2}
\end{align}
is a property of the initial state $\hat{\rho}$, the parameter-encoding Hamiltonian $\hat{H}$ and the observable $\hat{X}$.

The achievable sensitivity~(\ref{eq:mmoments2}) depends on the choice of the observable $\hat{X}$. This motivates the introduction of an optimal 
metrological squeezing parameter for a family of accessible operators $\hat{\mathbf{H}}=(\hat{H}_1,\dots,\hat{H}_K)$ as
\begin{align} \label{eq:xiQ}
\chi_{\rm opt}^2[\hat{\rho},\hat{H},\hat{\mathbf{H}}] := \min_{ \hat{X}\in\mathrm{span}(\hat{\mathbf{H}})} \chi^2[\hat{\rho},\hat{H},\hat{X}].
\end{align}
The analytical optimization over arbitrary linear combinations of accessible operators 
(namely $\hat{H} = \hat{H}_{\mathbf{n}}=\mathbf{n}\cdot\hat{\mathbf{H}}=\sum_{k=1}^Kn_k\hat{H}_k$ and 
$\hat{X}= \hat{H}_{\mathbf{m}}$ with $\mathbf{n},\mathbf{m}\in\mathbb{R}^K$) is one of the main results of this paper. For the inverse of Eq.~(\ref{eq:xiQ}), we obtain 
\begin{align}\label{eq:thm}
\chi_{\rm opt}^{-2}[\hat{\rho},\hat{H}_{\mathbf{n}},\hat{\mathbf{H}}]:=\max_{\mathbf{m}}\frac{|\langle [\hat{H}_{\mathbf{m}},\hat{H}_{\mathbf{n}}]\rangle_{\hat{\rho}}|^2}{(\Delta \hat{H}_{\mathbf{m}})_{\hat{\rho}}^2}=\mathbf{n}^T\mathbf{M}[\hat{\rho},\hat{\mathbf{H}}]\mathbf{n},
\end{align}
for all $\mathbf{n}$ and $\mathbf{m}$, where we introduced the moment matrix
\begin{align}\label{eq:momentmatrixNOX}
\mathbf{M}[\hat{\rho},\hat{\mathbf{H}}]=\mathbf{C}[\hat{\rho},\hat{\mathbf{H}}]^T\boldsymbol{\Gamma}[\hat{\rho},\hat{\mathbf{H}}]^{-1}\mathbf{C}[\hat{\rho},\hat{\mathbf{H}}].
\end{align}
Here, $\boldsymbol{\Gamma}[\hat{\rho},\hat{\mathbf{H}}]$ is the covariance matrix  with elements $(\boldsymbol{\Gamma}[\hat{\rho},\hat{\mathbf{H}}])_{kl}=\mathrm{Cov}(\hat{H}_k,\hat{H}_l)_{\hat{\rho}}$, which is symmetric $\boldsymbol{\Gamma}[\hat{\rho},\hat{\mathbf{H}}]=\boldsymbol{\Gamma}[\hat{\rho},\hat{\mathbf{H}}]^T$ and positive semidefinite for all $\hat{\mathbf{H}}$. We further assume $\boldsymbol{\Gamma}[\hat{\rho},\hat{\mathbf{H}}]$ to be positive definite and hence invertible, which excludes the situation where $\hat{\rho}$ has zero variance for some $\hat{H}_{\mathbf{n}}$. The real-valued, skew-symmetric commutator matrix $\mathbf{C}[\hat{\rho},\hat{\mathbf{H}}]=-\mathbf{C}[\hat{\rho},\hat{\mathbf{H}}]^T$ has elements $(\mathbf{C}[\hat{\rho},\hat{\mathbf{H}}])_{kl}=-i\langle[\hat{H}_k,\hat{H}_l]\rangle_{\hat{\rho}}$. The maximum in~(\ref{eq:thm}) is reached for 
\begin{align}\label{eq:optobs}
\mathbf{m}=\alpha\boldsymbol{\Gamma}[\hat{\rho},\hat{\mathbf{H}}]^{-1}\mathbf{C}[\hat{\rho},\hat{\mathbf{H}}]\mathbf{n},
\end{align}
where $\alpha\in\mathbb{R}$ is a normalization constant. 
To prove Eq.~(\ref{eq:thm}), we write $(\Delta \hat{H}_{\mathbf{m}})_{\hat{\rho}}^2=\mathbf{m}^T\boldsymbol{\Gamma}[\hat{\rho},\hat{\mathbf{H}}]\mathbf{m}$ and $\langle[\hat{H}_{\mathbf{m}},\hat{H}_{\mathbf{n}}]\rangle_{\hat{\rho}}=\sum_{kl}m_kn_l\langle[\hat{H}_k,\hat{H}_l]\rangle_{\hat{\rho}}=i\mathbf{m}^T\mathbf{C}[\hat{\rho},\hat{\mathbf{H}}]\mathbf{n}$. The Cauchy-Schwarz inequality $|\mathbf{u}^T\mathbf{v}|^2\leq (\mathbf{v}^T\mathbf{v})(\mathbf{u}^T\mathbf{u})$ holds for arbitrary vectors $\mathbf{u}$ and $\mathbf{v}$ and is saturated if and only if $\mathbf{u}=\alpha\mathbf{v}$ with some normalization constant $\alpha$. Inserting $\mathbf{u}=\boldsymbol{\Gamma}[\hat{\rho},\hat{\mathbf{H}}]^{\frac{1}{2}} \mathbf{m}$ and $\mathbf{v}=\boldsymbol{\Gamma}[\hat{\rho},\hat{\mathbf{H}}]^{-\frac{1}{2}}\mathbf{C}[\hat{\rho},\hat{\mathbf{H}}]\mathbf{n}$ yields the statement. 

The optimized squeezing coefficient~(\ref{eq:xiQ}) depends on the available set $\hat{\mathbf{H}}$ of accessible observables.
If $\hat{\mathbf{H}}$ contains all linear observables, Eq.~(\ref{eq:xiQ}) provides an analytically-optimized linear squeezing parameter.
By adding nonlinear observables, we can introduce metrological nonlinear squeezing parameters for non-Gaussian states, as we will discuss below. 
For all $\hat{X}$, $\hat{H}$ and $\hat{\mathbf{H}}$, the parameters $\chi^{-2}[\hat{\rho},\hat{H},\hat{X}]$ and $\chi_{\rm opt}^{-2}[\hat{\rho},\hat{H},\hat{\mathbf{H}}]$ are convex functions of $\hat{\rho}$, which ensures that they are maximized by pure states. Finally, maximizing over all possible observables $\hat{X}$, we have  that 
$\max_{ \hat{X}} \chi^{-2}[\hat{\rho},\hat{H},\hat{X}] = F_Q[\hat{\rho},\hat{H}]$ namely, the maximal squeezing parameter coincides with the quantum Fisher information~\cite{Supp}.

Besides determining the optimal measurement observable $\hat{H}_{\mathbf{m}}$, we may use Eq.~(\ref{eq:thm}) to find the evolution Hamiltonian $\hat{H}_{\mathbf{n}}$ that leads to the highest possible phase estimation sensitivity. This Hamiltonian is identified as $\hat{H}_{\mathbf{n}_{\max}}$, where $\mathbf{n}_{\max}$ is the maximum eigenvector of $\mathbf{M}[\hat{\rho},\hat{\mathbf{H}}]$ and the obtained sensitivity is the corresponding eigenvalue $\lambda_{\max}(\mathbf{M}[\hat{\rho},\hat{\mathbf{H}}])$. If $\mathbf{C}[\hat{\rho},\hat{\mathbf{H}}]\mathbf{n}_{\max}$ is a maximal eigenvector of $\boldsymbol{\Gamma}[\hat{\rho},\hat{\mathbf{H}}]^{-1}$, the optimal measurement $\mathbf{m}$ is achieved by $\mathbf{m}_{\max}=\alpha\boldsymbol{\Gamma}[\hat{\rho},\hat{\mathbf{H}}]^{-1}\mathbf{C}[\hat{\rho},\hat{\mathbf{H}}]\mathbf{n}_{\max}=\alpha'\mathbf{C}[\hat{\rho},\hat{\mathbf{H}}]\mathbf{n}_{\max}$, where we used the eigenvalue property and $\alpha,\alpha'$ are normalization constants. 

Finally, to quantify the achievable metrological sensitivity enhancement, we introduce the coefficient
\begin{align}\label{eq:xi}
\xi^{2}_{\rm opt}[\hat{\rho},\hat{H},\hat{\mathbf{H}}]:=\frac{F_{\mathrm{SN}}[\hat{H}]}{\chi_{\rm opt}^{-2}[\hat{\rho},\hat{H},\hat{\mathbf{H}}]},
\end{align}
where $F_{\mathrm{SN}}[\hat{H}]=\max_{\hat{\rho}_{\mathrm{cl}}}F_{Q}[\hat{\rho}_{\mathrm{cl}},\hat{H}]$ indicates the shot-noise (SN) limit, i.e., the maximal quantum Fisher information for classical states
namely particle-separable states in a many-spin system \cite{PS09} or coherent states in the continuous-variable regime \cite{MandelWolf,Luis}.

\begin{figure*}[tb]
\centering
\includegraphics[width=.99\textwidth]{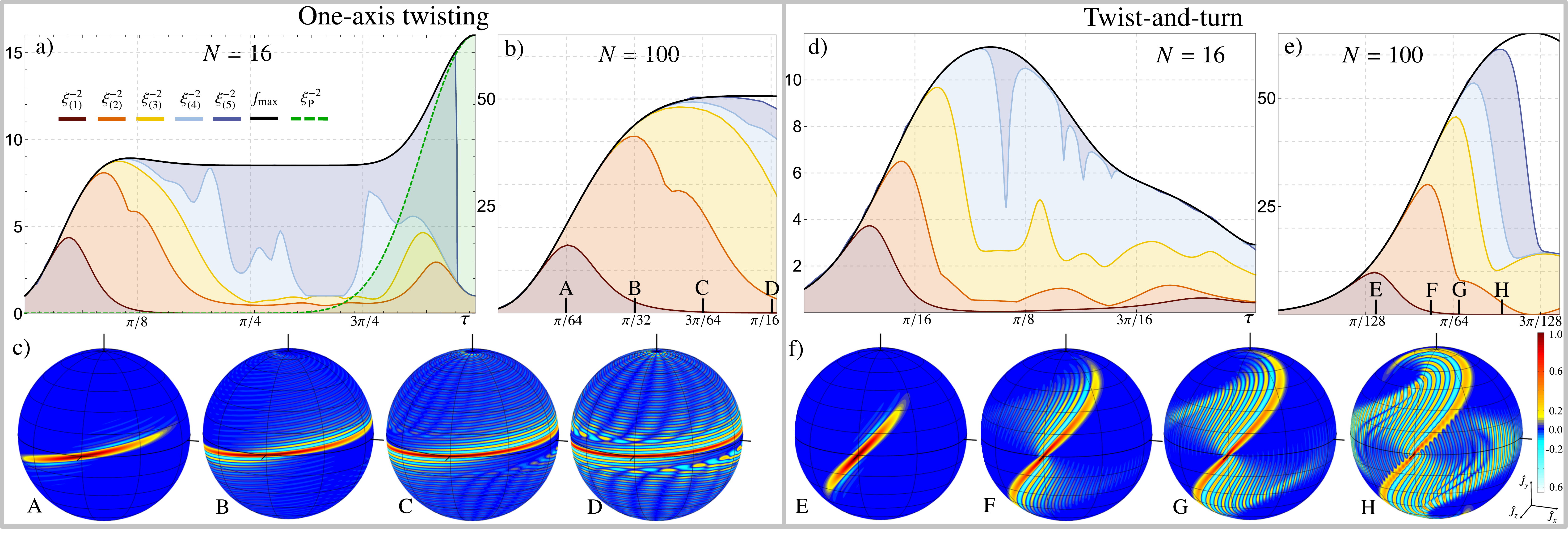}
\caption{Analytically optimized nonlinear spin squeezing coefficients $\xi^{-2}_{(K)}$ for $K=2,\dots,5$ compared to the linear spin squeezing coefficient $\xi^{-2}_{(1)}$, the spin parity squeezing coefficient $\xi^{-2}_{\mathrm{P}}$, and the quantum Fisher density $f_{\max}$ for the states $|\Psi_{\rm OAT}(\tau)\rangle$ [$N=16$ in a) and $N=100$ in b)], as well as the Wigner function of selected states with $N=100$ (c). Panels d), e) and f) show the same data for the states $|\Psi_{\rm TAT}(\tau)\rangle$.}
\label{fig:sqz}
\end{figure*}

\textit{Nonlinear spin squeezing coefficients.}---Let us consider the case of an $N$-qubit system described by collective spin operators $\hat{\mathbf{J}}=(\hat{J}_x,\hat{J}_y,\hat{J}_z)$ where $\hat{J}_\alpha=\sum_{i=1}^N\hat{\sigma}^{(i)}_\alpha/2$ and the $\hat{\sigma}^{(i)}_\alpha$ are the Pauli matrices for $\alpha=x,y,z$. The operators $\hat{\mathbf{J}}$ are linear in a sense that they do not involve spin-spin interactions. Particle-separable states $\hat{\rho}_{\mathrm{p-sep}}=\sum_{\gamma}p_{\gamma}\hat{\rho}^{(1)}_{\gamma}\otimes\cdots\otimes\hat{\rho}^{(N)}_{\gamma}$ can at most have a sensitivity of $F_{\mathrm{SN}}[\hat{J}_{\mathbf{n}}]=\max_{\hat{\rho}_{\mathrm{p-sep}}}F_Q[\hat{\rho}_{\mathrm{p-sep}},\hat{J}_{\mathbf{n}}]=N$ \cite{PS09}, where $p_{\gamma}$ describes a probability distribution and the $\hat{\rho}^{(k)}_{\gamma}$ are local states of the $k$th qubit. As the achievable quantum enhancement increases with the number of entangled particles, we obtain with Ref.~\cite{HyllusPRA2012} that $\xi^{-2}_{\rm opt}[\hat{\rho},\hat{J}_{\mathbf{n}},\hat{\mathbf{H}}]>k$ reveals multiparticle entanglement of at least $k$ qubits, where $\hat{J}_{\mathbf{n}}=\mathbf{n}\cdot\hat{\mathbf{J}}$ is an arbitrary collective spin operator with $\mathbf{n}\in\mathbb{R}^3$ and $|\mathbf{n}|^2=1$ and the elements of $\hat{\mathbf{H}}$ can be nonlinear \cite{note4}.

Using Eq.~(\ref{eq:xi}), we now introduce a fully optimized (linear) spin squeezing coefficient as $\xi_{(1)}^{2}[\hat{\rho}]=\min_{\mathbf{n}}\xi_{\rm opt}^{2}[\hat{\rho},\hat{J}_{\mathbf{n}},\hat{\mathbf{J}}]$. This is equivalent to the spin squeezing coefficient first introduced by Wineland \textit{et al.} \cite{Wineland}. Both directions for the measurement and the parameter-encoding evolution are optimized for sensitivity. The implied optimization problem can be involved (see, e.g., Ref.~\cite{Ma}). Using Eq.~(\ref{eq:thm}) and $\max_{\mathbf{n}}\chi_{\rm opt}^{-2}[\hat{\rho},\hat{J}_{\mathbf{n}},\hat{\mathbf{J}}]=\lambda_{\max}(\mathbf{M}[\hat{\rho},\hat{\mathbf{J}}])$, we obtain
\begin{align}\label{eq:spinsqz}
\xi_{(1)}^{2}[\hat{\rho}]=\min_{\mathbf{m},\mathbf{n}}\frac{N(\Delta\hat{J}_{\mathbf{m}})^2_{\hat{\rho}}}{\vert \langle [\hat{J}_{\mathbf{m}},\hat{J}_{\mathbf{n}}] \rangle_{\hat{\rho}} \vert^2}=\frac{N}{\lambda_{\max}(\mathbf{M}[\hat{\rho},\hat{\mathbf{J}}])},
\end{align}
which solves this problem analytically for arbitrary states $\hat{\rho}$ without constraints. A state $\hat{\rho}$ is spin squeezed if $\xi_{(1)}^{2}[\hat{\rho}] < 1$. 

To assess the sensitivity of non-Gaussian spin states we define optimized nonlinear spin-squeezing coefficients of order $K$ as $\xi_{(K)}^{2}[\hat{\rho}]=\min_{\mathbf{n}}\xi_{\rm opt}^{2}[\hat{\rho},\hat{J}_{\mathbf{n}},\hat{\mathbf{J}}^{(K)}]$. The measurement observable is optimized for sensitivity over families of operators $\hat{\mathbf{J}}^{(K)}$ that include, beyond the linear operators $\hat{\mathbf{J}}$, also symmetric products of up to $K$ linear operators. For example, the second-order spin-squeezing coefficient is obtained from Eq.~(\ref{eq:thm}) as
\begin{align}\label{eq:xi2}
\xi^{2}_{(2)}[\hat{\rho}]=\min_{\mathbf{n}\in\mathbb{R}^3}\min_{\mathbf{m}\in\mathbb{R}^9}\frac{N(\Delta \hat{J}^{(2)}_{\mathbf{m}})_{\hat{\rho}}^2}{|\langle [\hat{J}^{(2)}_{\mathbf{m}},\hat{J}_{\mathbf{n}}]\rangle_{\hat{\rho}}|^2}=\frac{N}{\lambda_{\max}(\tilde{\mathbf{M}}[\hat{\rho},\hat{\mathbf{J}}^{(2)}])},
\end{align}
where $\hat{\mathbf{J}}^{(2)}=(\hat{J}_x,\hat{J}_y,\hat{J}_z,\hat{J}_x^2,\hat{J}_y^2,\hat{J}_y^2,\frac{1}{2}\{\hat{J}_x,\hat{J}_y\},\frac{1}{2}\{\hat{J}_x,\hat{J}_z\},\frac{1}{2}\{\hat{J}_y,\hat{J}_z\})$ contains all linear and symmetric quadratic collective spin operators and $\hat{J}^{(2)}_{\mathbf{m}}=\mathbf{m}\cdot\hat{\mathbf{J}}^{(2)}$~\cite{note5}. The optimization problem for $\mathbf{m}$ is solved using Eq.~(\ref{eq:thm}). Note that we always assume a linear generator $\hat{H}=\hat{J}_{\mathbf{n}}$ for the phase imprinting evolution. Thus, for the optimization of $\mathbf{n}$, we restrict to the first three rows and columns of $\mathbf{M}[\hat{\rho},\hat{\mathbf{J}}^{(2)}]$, giving rise to the principal submatrix $\tilde{\mathbf{M}}[\hat{\rho},\hat{\mathbf{J}}^{(2)}]$ \cite{Supp}. The higher-order coefficients $\xi^{2}_{(K)}[\hat{\rho}]$ are obtained analogously. The procedure is experimentally challenging for large values of $K$ but less costly than full quantum state tomography. This defines a hierarchy $\xi^{-2}_{(1)}[\hat{\rho}] \leq \xi^{-2}_{(2)}[\hat{\rho}] \leq \xi^{-2}_{(3)}[\hat{\rho}]\leq ... $ that generalizes and improves the linear spin squeezing coefficient, capturing a larger and larger set of metrologically useful states.

\textit{Collective spin systems: Nonlinear evolution.}---Benchmark examples of non-Gaussian spin states are obtained by the one-axis-twisting (OAT) \cite{KitagawaUeda} evolution $|\Psi_{\rm OAT}(\tau)\rangle=e^{-i \tau \hat{J}_y^2}|\Psi(0)\rangle$ 
starting from a coherent spin state pointing in the $z$ direction, i.e., $|\Psi(0)\rangle=|N/2,N/2\rangle_z$ is a simultaneous eigenstate of $\hat{\mathbf{J}}^2$ and $\hat{J}_z$ and here we assume $N$ to be even. For short times, the evolution generates spin-squeezed states, as captured by the linear spin-squeezing coefficient~\cite{KitagawaUeda,RMP}.
For $\tau \gtrsim 1/\sqrt{N}$ spin squeezing is lost as the state wraps around the Bloch sphere and becomes non-Gaussian \cite{PS09}, generating a GHZ state at $\tau=\pi/2$ \cite{MolmerPRL1999} and a full revival of the initial spin coherent state at $\tau=\pi$. 
This dynamics has been realized on short time scales~\cite{RMP} with large ensembles of Bose-Einstein condensates \cite{Riedel}, ultracold atoms in a cavity system~\cite{LerouxPRL2010}, or trapped ions \cite{Bohnet}, and on longer times in experiments with smaller numbers of trapped ions~\cite{MonzPRL2011} as well as, recently, for the electronic spin $J=8$ of dysprosium atoms~\cite{ChalopinArXiv}. Lower bounds on the Fisher information for the states $|\Psi_{\rm OAT}(\tau)\rangle$ have been studied numerically in Ref.~\cite{Frowis2015}. In Fig.~\ref{fig:sqz} we show the analytically optimized linear and nonlinear spin-squeezing coefficients $\xi^{-2}_{(K)}[|\Psi_{\mathrm{OAT}}(\tau)\rangle]$ for $K=1,\dots,5$ as a function of the evolution time, 
compared to the maximal quantum Fisher density $f_{\max}[|\Psi_{\mathrm{OAT}}(\tau)\rangle]=\max_{\mathbf{n}}F_Q[|\Psi(\tau)\rangle,\hat{J}_{\mathbf{n}}]/N$.
For $N=16$ the metrological sensitivity over almost the entire evolution period is captured by $\xi^{-2}_{(K)}[\hat{\rho}]$ with $K\leq 5$ [Fig.~\ref{fig:sqz} a)]. 
For $N=100$, sensitivities up to $N^2/2$ are still revealed [Fig.~\ref{fig:sqz} b)]
and are achieved by highly non-Gaussian states [Fig.~\ref{fig:sqz} c)].
For long evolution times, the characterization is complemented by the spin-parity-squeezing coefficient, $\xi^{2}_{\mathrm{P}}[\hat{\rho}]=\xi^{2}[\hat{\rho},\hat{J}_z,\hat{P}]$, where $\hat{P}=(-1)^{J-\hat{J}_x}$. The coefficient $\xi^{2}_{\mathrm{P}}[\hat{\rho}]$ is particularly suitable in the vicinity of the GHZ state~\cite{BollingerPRA1996} at $\tau=\pi/2$, as is shown by the green dashed line in Fig.~\ref{fig:sqz} a) \cite{footnoteParity}.

A faster generation of entanglement is possible by the so-called twist-and-turn evolution $|\Psi_{\rm TAT}(\tau)\rangle=e^{-i \tau (\hat{J}_y^2-\frac{N}{2}\hat{J}_z)}|\Psi(0)\rangle$ \cite{Strobel,Muessel,Sorelli}. 
Also in this case, whereas the onset of non-Gaussianity produces a rapid decay of the linear squeezing coefficient ($K=1$), nonlinear squeezing coefficients of moderate order are sufficient to capture large sensitivities and reveal significant amounts of particle entanglement 
at experimentally-relevant short times [Fig.~\ref{fig:sqz} d) and e)]. The Wigner functions~\cite{RMP} [Fig.~\ref{fig:sqz} c) and f)] reflect the non-Gaussian nature of the generated states revealed by these methods.

\textit{Nonlinear continuos-variable squeezing parameters and Fock-state sensing.}---In the continuous variable case, the nonlinear squeezing coefficient is defined from higher-order combinations of 
$\hat{x}=(\hat{a}+\hat{a}^{\dagger})/\sqrt{2}$ and $\hat{p}=i(\hat{a}^{\dagger}-\hat{a})/\sqrt{2}$, i.e., phase space quadrature operators 
for a bosonic single-mode field with annihilation operator $\hat{a}$ \cite{note6}. A particularly important application is the sensing of displacement amplitudes, which can be used to estimate small forces and fields \cite{Wolf}. Displacements are generated by linear combinations of $\hat{x}$ and $\hat{p}$, i.e., $\hat{D}(\alpha)=\exp(\alpha \hat{a}^{\dagger}-\alpha^* \hat{a})=\exp\left(i\hat{q}_{\mathbf{n}}\theta\right)$. Here the amplitude of $\alpha=\theta e^{-i\phi}/\sqrt{2}$ characterizes the phase parameter $\theta$ of interest and its phase $\phi$ determines the ``direction" of the displacement via the quadrature $\hat{q}_{\mathbf{n}}=n_1\hat{x}+n_2\hat{p}$ with $n_1=\sin(\phi)$ and $n_2=-\cos(\phi)$. 
Shot-noise sensitivity is attained by coherent states $|\alpha\rangle=\hat{D}(\alpha)|0\rangle$, generated from the vacuum $|0\rangle$. The displacement sensitivity of $|\alpha\rangle$ is independent of $\alpha$ and we find $F_\mathrm{SN}[\hat{q}_{\mathbf{n}}]=\max_{|\alpha\rangle;|\alpha|^2=N}F_Q[|\alpha\rangle,\hat{q}_{\mathbf{n}}]=F_Q[|0\rangle,\hat{q}_{\mathbf{n}}]=2$.

Fock states $|N\rangle=(\hat{a}^{\dagger})^N/\sqrt{N!}|0\rangle$ have particularly appealing properties for displacement amplitude sensing. They can be generated in a variety of quantum systems, e.g., by manipulating the motion of trapped ions \cite{Leibfried}, by strong atom-cavity interactions \cite{DelegliseNATURE2008}, or by optical nonlinear \cite{Sciarrino} or superradiant processes \cite{Paulisch}.
Due to their isotropic concentric fringes in phase space, Fock states yield sub-shot-noise sensitivity for any displacement generated by $\hat{q}_{\mathbf{n}}$. Fock states lead to $F_Q[|N\rangle,\hat{q}_{\mathbf{n}}]=4N+2$, which indicates a quantum enhancement for all $N>0$, independently of $\mathbf{n}$.

As Fock states are non-Gaussian, their characteristics cannot be sufficiently uncovered by measuring linear observables. In fact, even second-order observables are insufficient \cite{Supp}. Let us therefore extend the family of accessible operators by adding four nonlinear observables of third order: $
\hat{\mathbf{H}}=(\hat{x},\hat{p},\hat{x}^3, \frac{\hat{p}\hat{x}^2 + \hat{x}\hat{p}\hat{x}+ \hat{x}^2\hat{p}}{3},\frac{\hat{x}\hat{p}^2+\hat{p}\hat{x}\hat{p}+\hat{p}^2\hat{x}}{3},\hat{p}^3)$. We now find with Eq.~(\ref{eq:thm})
\begin{align}\label{eq:9}
\chi_{\rm opt}^{-2}[|N\rangle,\hat{q}_{\mathbf{n}},\hat{\mathbf{H}}]=4N + 2,
\end{align}
which coincides with the states' quantum Fisher information, i.e., the full metrological potential. The optimal observable $\hat{H}_{\mathbf{m}_{\mathrm{opt}}}=\mathbf{m}_{\mathrm{opt}}\cdot\hat{\mathbf{H}}$ for displacement sensing with a Fock state $|N\rangle$ is given according to Eq.~(\ref{eq:optobs}) by $
\mathbf{m}_{\mathrm{opt}}=c_N(-(2 N+1) n_2, (2 N+1)n_1, n_2, -n_1, n_2, -n_1)$ with the normalization constant $c_N=1/\sqrt{3 + 4 N (N+1)}$ and $n_1,n_2$ characterize the generating quadrature $\hat{q}_{\mathbf{n}}$.

\textit{Conclusions.}---We have extended the concept of metrological squeezing to arbitrary nonlinear observables of discrete- 
and continuous variables by devising a systematic optimization of the measurement observable for quantum metrology. 
Our methods provide the smallest achievable phase uncertainty for any given set of accessible measurement observables. 
This generalizes the standard (linear) squeezing parameters that are obtained by measuring only linear observables. 
Nonlinear squeezing relates the metrological quantum enhancement of non-Gaussian states to the squeezed 
variance of an optimal nonlinear observable, which is identified from the accessible set. 
This paves the way for implementable strategies to characterize metrological sensitivity and entanglement of non-Gaussian states, and to harness their increased potential in quantum metrology experiments. Our methods can be readily applied to non-Gaussian states of quantum light, as well as oversqueezed spin states of cold atoms or trapped ions.

\textit{Acknowledgments.---}M.G. acknowledges funding by the Alexander von Humboldt foundation. 
This work has been supported by the European Commission through the QuantERA ERA-NET Cofund in Quantum Technologies projects ``Q-Clocks'' and ``CEBBEC'', 
by the EURAMET Empir project ``USOQS'', and by  
the LabEx ENS-ICFP:ANR-10-LABX-0010/ANR-10-IDEX-0001-02 PSL*. M.G. thanks M. Walschaers for useful discussions.

\section{Proof of the convexity property}
For $\hat{\rho}=\sum_{\gamma}p_{\gamma}\hat{\rho}_{\gamma}$, concavity of the variance implies that $\chi^{-2}[\hat{\rho},\hat{H},\hat{X}]\leq\left|\sum_{\gamma}p_{\gamma}\langle[\hat{X},\hat{H}]\rangle_{\hat{\rho}_{\gamma}}\right|^2/\left(\sum_{\gamma}p_{\gamma}(\Delta\hat{X})^2_{\hat{\rho}_{\gamma}}\right)$. From the Cauchy-Schwarz inequality with $u_{\gamma}=\sqrt{p_{\gamma}}(\Delta\hat{X})_{\hat{\rho}_{\gamma}}$ and $v_{\gamma}=\sqrt{p_{\gamma}}\langle[\hat{X},\hat{H}]\rangle_{\hat{\rho}_{\gamma}}/(\Delta\hat{X})_{\hat{\rho}_{\gamma}}$ follows that $\chi^{-2}[\hat{\rho},\hat{H},\hat{X}]\leq \sum_{\gamma}p_{\gamma}\left|\langle[\hat{X},\hat{H}]\rangle_{\hat{\rho}_{\gamma}}\right|^2/(\Delta\hat{X})^2_{\hat{\rho}_{\gamma}}=\sum_{\gamma}p_{\gamma}\chi^{-2}[\hat{\rho}_{\gamma},\hat{H},\hat{X}]$, which demonstrates the convexity of $\chi^{-2}[\hat{\rho},\hat{H},\hat{X}]$. Convexity of $\chi_{\rm opt}^{-2}[\hat{\rho},\hat{H},\hat{\mathbf{H}}]$ then follows from $\chi_{\rm opt}^{-2}[\hat{\rho},\hat{H},\hat{\mathbf{H}}]=\chi^{-2}[\hat{\rho},\hat{H},\hat{X}_{\mathrm{opt}}]\leq\sum_{\gamma}p_{\gamma}\chi^{-2}[\hat{\rho}_{\gamma},\hat{H},\hat{X}_{\mathrm{opt}}]\leq\sum_{\gamma}p_{\gamma}\max_{\hat{X}\in\mathrm{span}(\hat{\mathbf{H}})}\chi^{-2}[\hat{\rho}_{\gamma},\hat{H},\hat{X}]=\sum_{\gamma}p_{\gamma}\chi_{\mathrm{opt}}^{-2}[\hat{\rho}_{\gamma},\hat{H},\hat{\mathbf{H}}]$.

\section{Saturating the quantum Fisher information}
Given an orthonormal set of projectors $\hat{\boldsymbol{\Pi}}=\{\hat{\Pi}_x\}$, equivalence of the squeezing parameter~(\ref{eq:mmoments2}) and the Fisher information, $\chi^{-2}[\hat{\rho}(\theta),\hat{X}]=F[\hat{\rho}(\theta),\hat{X}]$, is established, e.g., for $\hat{X}=\sum_x\left(\langle\hat{X}\rangle_{\hat{\rho}(\theta)}+\frac{\partial}{\partial \theta}\log p(x|\theta)\right)\hat{\Pi}_x$~\cite{Kholevo,Frowis2015}. The squeezing parameter thus equals the quantum Fisher information when $\hat{\boldsymbol{\Pi}}$ is the set of projectors that yields equality between the classical and quantum Fisher information, e.g., the projectors onto the eigenstates of the symmetric logarithmic derivative \cite{BraunsteinPRL1994}. Inserting the spectral decomposition $\hat{L}=\sum_x\lambda_x\hat{\Pi}_x$ for a fixed value of $\theta$ yields $\frac{\partial}{\partial \theta}\log p(x|\theta)=\lambda_x$ and $\langle\hat{L}\rangle_{\hat{\rho}(\theta)}=0$, leading to the optimal measurement operator $\hat{X}=\hat{L}$. We may explicitly verify the equality $\chi^{-2}[\hat{\rho}(\theta),\hat{L}] = (\Delta\hat{L})^{-2}_{\hat{\rho}(\theta)}(\frac{d\langle \hat{L}\rangle_{\hat{\rho}(\theta)}}{d\theta})^{2}=\mathrm{Tr}\{\hat{\rho}(\theta)\hat{L}^2\}=F_Q[\hat{\rho}(\theta)]$.

\section{Nonlinear spin squeezing coefficients}
The nonlinear spin squeezing coefficients are defined as
\begin{align}\label{eq:xi2}
\xi^{2}_{(K)}[\hat{\rho}]=\min_{\mathbf{n}\in\mathbb{R}^3}\min_{\mathbf{m}}\frac{N(\Delta \hat{J}^{(K)}_{\mathbf{m}})_{\hat{\rho}}^2}{|\langle [\hat{J}^{(K)}_{\mathbf{m}},\hat{J}_{\mathbf{n}}]\rangle_{\hat{\rho}}|^2},
\end{align}
where $K$ is the order of the linearity. To solve the optimization problem, we make use of Eq.~(\ref{eq:thm}). Thus, we determine the matrix elements $(\boldsymbol{\Gamma}[\hat{\rho},\hat{\mathbf{J}}^{(K)}])_{kl}=\frac{1}{2}\langle\hat{J}^{(K)}_k\hat{J}^{(K)}_l+\hat{J}^{(K)}_l\hat{J}^{(K)}_k\rangle_{\hat{\rho}}-\langle\hat{J}^{(K)}_k\rangle_{\hat{\rho}}\langle\hat{J}^{(K)}_l\rangle_{\hat{\rho}}$ and $(\mathbf{C}[\hat{\rho},\hat{\mathbf{J}}^{(K)}])_{kl}=-i\langle[\hat{J}^{(K)}_k,\hat{J}^{(K)}_l]\rangle_{\hat{\rho}}$, where $\hat{\mathbf{J}}^{(K)}$ is the vector of nonlinear spin operators up to order $K$. 

In general, the number of elements in the set $\hat{\mathbf{J}}^{(K)}$ scales exponentially with $K$, 
i.e., all unique products of up to $K$ elements of $\hat{\mathbf{J}}$ (rendered Hermitian by adding the same product in reverse order and dividing by two), whereas the cost of full state tomography without assumptions grows exponentially with $N$. For the states considered in this manuscript, restricting to fully symmetric products is sufficient. That is, instead of treating, e.g., $(\hat{J}_x\hat{J}_y\hat{J}_y+\hat{J}_y\hat{J}_y\hat{J}_x)/2$ and $\hat{J}_y\hat{J}_x\hat{J}_y$ as two independent observables, we only consider fully symmetric products of the kind $(\hat{J}_x\hat{J}_y\hat{J}_y+\hat{J}_y\hat{J}_x\hat{J}_y+\hat{J}_y\hat{J}_y\hat{J}_x)/3$, etc., which reduces the effective number of elements to $(K + 1) (K + 2)/2$ and a more favorable quadratic scaling with $K$. For example, for $K=3$, we have
\begin{align}
&\hat{\mathbf{J}}^{(3)}=\left(\hat{J}_x,\hat{J}_y,\hat{J}_z,\hat{J}_x^2,\hat{J}_y^2,\hat{J}_y^2,\frac{1}{2}\{\hat{J}_x,\hat{J}_y\},\frac{1}{2}\{\hat{J}_x,\hat{J}_z\},\frac{1}{2}\{\hat{J}_y,\hat{J}_z\},\right.\notag\\&\left.\hat{J}_x^3,\frac{1}{3}\left[\hat{J}_x^2\hat{J}_y+\hat{J}_x\hat{J}_y\hat{J}_x+\hat{J}_y\hat{J}_x^2\right],\frac{1}{3}\left[\hat{J}_x^2\hat{J}_z+ \hat{J}_x\hat{J}_z\hat{J}_x+\hat{J}_z\hat{J}_x^2\right],\right.\notag\\&\left.\frac{1}{3}\left[\hat{J}_y^2\hat{J}_x+\hat{J}_y\hat{J}_x\hat{J}_y+\hat{J}_x\hat{J}_y^2\right],\hat{J}_y^3,\frac{1}{3}\left[\hat{J}_y^2\hat{J}_z+\hat{J}_y\hat{J}_z\hat{J}_y+\hat{J}_z\hat{J}_y^2\right] ,\right.\notag\\&\left.\frac{1}{3}\left[\hat{J}_z^2\hat{J}_x+\hat{J}_z\hat{J}_x\hat{J}_z+\hat{J}_x\hat{J}_z^2\right],\frac{1}{3}\left[\hat{J}_z^2\hat{J}_y+\hat{J}_z\hat{J}_y\hat{J}_z+\hat{J}_y\hat{J}_z^2\right],\hat{J}_z^3,\right.\notag\\&\left.\frac{1}{6}\left[\hat{J}_x\hat{J}_y\hat{J}_z+ \hat{J}_x\hat{J}_z\hat{J}_y+\hat{J}_y\hat{J}_x\hat{J}_z+\hat{J}_y\hat{J}_z\hat{J}_x+\hat{J}_z\hat{J}_x\hat{J}_y+\hat{J}_z\hat{J}_y\hat{J}_x\right]\right),\notag
\end{align}
where $\{\hat{A},\hat{B}\}=\hat{A}\hat{B}+\hat{B}\hat{A}$ is the anti-commutator. This vector contains $3$ linear operators (the first three elements), $6$ second-order and $10$ third-order nonlinear spin operators. Thus, $\boldsymbol{\Gamma}[\hat{\rho},\hat{\mathbf{J}}^{(3)}]$ and $\mathbf{C}[\hat{\rho},\hat{\mathbf{J}}^{(3)}]$ are $19\times 19$ matrices in this case. The procedure for higher orders is analogous. From these matrices, we determine the moment matrix $\mathbf{M}[\hat{\rho},\hat{\mathbf{J}}^{(K)}]=\mathbf{C}[\hat{\rho},\hat{\mathbf{J}}^{(K)}]^T\boldsymbol{\Gamma}[\hat{\rho},\hat{\mathbf{J}}^{(K)}]^{-1}\mathbf{C}[\hat{\rho},\hat{\mathbf{J}}^{(K)}]$, and with Eq.~(\ref{eq:thm}), we solve the minimization over $\mathbf{m}$ as
\begin{align}
\xi^{2}_{(K)}[\hat{\rho}]=\min_{\mathbf{n}\in\mathbb{R}^3}\frac{N}{\mathbf{n}^T\mathbf{M}[\hat{\rho},\hat{\mathbf{J}}^{(K)}]\mathbf{n}}.
\end{align}
To finally optimize over the orientation $\mathbf{n}\in\mathbb{R}^3$ of the linear phase-encoding transformations generated by $\hat{J}_{\mathbf{n}}$, we recall that the first three elements of $\hat{\mathbf{J}}^{(K)}$ are the three linear spin operators $\hat{J}_x$, $\hat{J}_y$, and $\hat{J}_z$. The achievable sensitivity for linear evolutions is therefore contained in the principal submatrix that is obtained by keeping only the first three rows and columns of the matrix $\mathbf{M}[\hat{\rho},\hat{\mathbf{J}}^{(K)}]$. Denoting this submatrix as $\tilde{\mathbf{M}}[\hat{\rho},\hat{\mathbf{J}}^{(K)}]$, the nonlinear spin squeezing coefficient is finally given as
\begin{align}
\xi^{2}_{(K)}[\hat{\rho}]=\frac{N}{\lambda_{\max}(\tilde{\mathbf{M}}[\hat{\rho},\hat{\mathbf{J}}^{(K)}])},
\end{align}
where $\lambda_{\max}$ denotes the largest eigenvalue.

\section{Nonlinear squeezing of Fock states}
\subsection{Insufficiency of second-order squeezing}
Let us first show that second-order nonlinear squeezing is insufficient to capture the sensitivity of Fock states. To this end, notice that for $\hat{\mathbf{H}}^{(2)}=(\hat{x},\hat{p},\hat{x}^2,\frac{1}{2}\{\hat{x},\hat{p}\},\hat{p}^2)$, the covariance and commutator matrices both attain a block-diagonal form
\begin{align}\label{eq:GammaFockBlockD}
\boldsymbol{\Gamma}[|N\rangle,\hat{\mathbf{H}}^{(2)}]=\begin{pmatrix}
\boldsymbol{\Gamma}[|N\rangle,\hat{\mathbf{r}}] & \mathbf{0}\\
\mathbf{0} & \boldsymbol{\Gamma}[|N\rangle,\hat{\mathbf{r}}^{(2)}] 
\end{pmatrix},
\end{align}
and
\begin{align}\label{eq:CFockBlockD}
\mathbf{C}[|N\rangle,\hat{\mathbf{H}}^{(2)}]=\begin{pmatrix}
\mathbf{C}[|N\rangle,\hat{\mathbf{r}}] & \mathbf{0}\\
\mathbf{0} & \mathbf{C}[|N\rangle,\hat{\mathbf{r}}^{(2)}]
\end{pmatrix},
\end{align}
where we have separated the components of $\hat{\mathbf{H}}^{(2)}$ into linear $\hat{\mathbf{r}}=(\hat{x},\hat{p})$ and quadratic contributions $\hat{\mathbf{r}}^{(2)}=(\hat{x}^2,\frac{1}{2}\{\hat{x},\hat{p}\},\hat{p}^2)$. The absence of off-diagonal terms is due to vanishing odd moments for Fock states. The second-order squeezing parameter is obtained by maximizing the quantity
\begin{align}\label{eq:sqzparam}
\chi^{-2}[|N\rangle,\hat{q}_{\mathbf{n}},\hat{H}^{(2)}_{\mathbf{m}}]&=\frac{\vert \langle [\hat{H}^{(2)}_{\mathbf{m}},\hat{q}_{\mathbf{n}}] \rangle_{|N\rangle} \vert^2}{(\Delta\hat{H}^{(2)}_{\mathbf{m}})^2_{|N\rangle}}
\end{align}
over $\mathbf{m}\in\mathbb{R}^5$ and $\mathbf{n}\in\mathbb{R}^2$, where $\hat{q}_{\mathbf{n}}=\mathbf{n}\cdot\hat{\mathbf{r}}$ and $\hat{H}^{(2)}_{\mathbf{m}}=\mathbf{m}\cdot\hat{\mathbf{H}}^{(2)}$. Now, we write
\begin{align}
(\Delta\hat{H}^{(2)}_{\mathbf{m}})^2_{|N\rangle}&=\mathbf{m}^T\boldsymbol{\Gamma}[|N\rangle,\hat{\mathbf{H}}^{(2)}]\mathbf{m}\notag\\
&=\mathbf{m}^{(1)T}\boldsymbol{\Gamma}[|N\rangle,\hat{\mathbf{r}}]\mathbf{m}^{(1)}+\mathbf{m}^{(2)T}\boldsymbol{\Gamma}[|N\rangle,\hat{\mathbf{r}}^{(2)}]\mathbf{m}^{(2)},
\end{align}
due to Eq.~(\ref{eq:GammaFockBlockD}), and we separated linear and quadratic coefficients of $\mathbf{m}=(m_1,m_2,m_3,m_4,m_5)$ as $\mathbf{m}^{(1)}=(m_1,m_2)$ and $\mathbf{m}^{(2)}=(m_3,m_4,m_5)$.
Similarly, we obtain
\begin{align}
\vert \langle [\hat{H}^{(2)}_{\mathbf{m}},\hat{q}_{\mathbf{n}}] \rangle_{|N\rangle} \vert^2&=\vert\mathbf{m}^T\mathbf{C}[|N\rangle,\hat{\mathbf{H}}^{(2)}]\begin{pmatrix}n_1\\n_2\\0\\0\\0\end{pmatrix}\vert^2\notag\\
&=\vert\mathbf{m}^{(1)T}\mathbf{C}[|N\rangle,\hat{\mathbf{r}}]\mathbf{n}\vert^2,
\end{align}
where we made use of Eq.~(\ref{eq:CFockBlockD}). Inserting these expressions into the squeezing parameter~(\ref{eq:sqzparam}), we obtain
\begin{align}
\chi^{-2}[|N\rangle,\hat{q}_{\mathbf{n}},\hat{H}^{(2)}_{\mathbf{m}}]&=\frac{\vert\mathbf{m}^{(1)T}\mathbf{C}[|N\rangle,\hat{\mathbf{r}}]\mathbf{n}\vert^2}{\mathbf{m}^{(1)T}\boldsymbol{\Gamma}[|N\rangle,\hat{\mathbf{r}}]\mathbf{m}^{(1)}+\mathbf{m}^{(2)T}\boldsymbol{\Gamma}[|N\rangle,\hat{\mathbf{r}}^{(2)}]\mathbf{m}^{(2)}}\notag\\
&\leq \frac{\vert\mathbf{m}^{(1)T}\mathbf{C}[|N\rangle,\hat{\mathbf{r}}]\mathbf{n}\vert^2}{\mathbf{m}^{(1)T}\boldsymbol{\Gamma}[|N\rangle,\hat{\mathbf{r}}]\mathbf{m}^{(1)}}=\chi^{-2}[|N\rangle,\hat{q}_{\mathbf{n}},\hat{q}_{\mathbf{m}^{(1)}}],\notag
\end{align}
since $\mathbf{m}^{(2)T}\boldsymbol{\Gamma}[|N\rangle,\hat{\mathbf{r}}^{(2)}]\mathbf{m}^{(2)}\geq 0$. Hence, the second-order squeezing parameter can never be larger than the one obtained from the first order. We must therefore include operators of at least third order to capture the sensitivity of Fock states under displacements. Indeed, as we show below, the full sensitivity can be revealed by considering only accessible operators of first and third order.

\subsection{Optimality of third-order squeezing}
Following the argumentation of the previous section, it is straightforward to see that measurements of second-order operators are not useful to enhance the sensitivity in combination with first- and third-order operators, and will therefore be omitted in the following. We thus consider the family of accessible operators $
\hat{\mathbf{H}}=(\hat{x},\hat{p},\hat{x}^3, \frac{\hat{p}\hat{x}^2 + \hat{x}\hat{p}\hat{x}+ \hat{x}^2\hat{p}}{3},\frac{\hat{x}\hat{p}^2+\hat{p}\hat{x}\hat{p}+\hat{p}^2\hat{x}}{3},\hat{p}^3)$, which lead to the covariance matrix
\begin{widetext}
\begin{align}
\boldsymbol{\Gamma}[|N\rangle,\hat{\mathbf{H}}]=\begin{pmatrix} N+\frac{1}{2} & 0 & \frac{3}{4} f_N(1) & 0 & \frac{1}{4} f_N(1) & 0 \\
 0 & N+\frac{1}{2} & 0 & \frac{1}{4} f_N(1) & 0 & \frac{3}{4} f_N(1) \\
 \frac{3}{4} f_N(1) & 0 & \frac{5}{8} (2 N (N (2 N+3)+4)+3) & 0 & \frac{1}{8} (2 N+1) f_N(-3) & 0 \\
 0 & \frac{1}{4} f_N(1) & 0 & \frac{1}{8} (2 N+1) f_N(7) & 0 & \frac{1}{8} (2 N+1) f_N(-3) \\
 \frac{1}{4} f_N(1) & 0 & \frac{1}{8} (2 N+1) f_N(-3) & 0 & \frac{1}{8} (2 N+1) f_N(7) & 0 \\
 0 & \frac{3}{4} f_N(1) & 0 & \frac{1}{8} (2 N+1) f_N(-3) & 0 & \frac{5}{8} (2 N (N (2 N+3)+4)+3)
\end{pmatrix},\notag
\end{align}
with $f_N(k)=2 N (N+1)+k$. Similarly, the commutator matrix for a Fock state yields
\begin{align}
\mathbf{C}[|N\rangle,\hat{\mathbf{H}}]=\begin{pmatrix} 
 0 & 1 & 0 & \frac{1}{2} (2 N+1) & 0 & \frac{3}{2} (2 N+1) \\
 -1 & 0 & -\frac{3}{2} (2 N+1) & 0 & -\frac{1}{2} (2 N+1) & 0 \\
 0 & \frac{3}{2} (2 N+1) & 0 & \frac{9}{4} (2 N (N+1)+1) & 0 & \frac{3}{4} (6 N (N+1)+1) \\
 -\frac{1}{2} (2 N+1) & 0 & -\frac{9}{4} (2 N (N+1)+1) & 0 & \frac{1}{4} (6 N (N+1)+5) & 0 \\
 0 & \frac{1}{2} (2 N+1) & 0 & -\frac{1}{4} (6 N (N+1)+5) & 0 & \frac{9}{4} (2 N (N+1)+1) \\
 -\frac{3}{2} (2 N+1) & 0 & -\frac{3}{4} (6 N (N+1)+1) & 0 & -\frac{9}{4} (2 N (N+1)+1) & 0
\end{pmatrix}.\notag
\end{align}
Using this in Eq.~(4) of the main manuscript yields the moment matrix
\begin{align}
\mathbf{M}[|N\rangle,\hat{\mathbf{H}}]=\begin{pmatrix} 
 4 N+2 & 0 & 6 N (N+1)+3 & 0 & 2 N (N+1)+1 & 0 \\
 0 & 4 N+2 & 0 & 2 N (N+1)+1 & 0 & 6 N (N+1)+3 \\
 6 N (N+1)+3 & 0 & 9g_N(19,43,10) & 0 & 3g_N(1,-5,-6) & 0 \\
 0 & 2 N (N+1)+1 & 0 & g_N(55,139,42) & 0 & 3g_N(1,-5,-6) \\
 2 N (N+1)+1 & 0 & 3g_N(1,-5,-6) & 0 & g_N(55,139,42) & 0 \\
 0 & 6 N (N+1)+3 & 0 & 3g_N(1,-5,-6) & 0 & 9g_N(19,43,10)
\end{pmatrix},\notag
\end{align}
\end{widetext}
with $g_N(k,l,m)=\frac{(N (N+1) (2 N (N+1) (2 N (N+1)+k)+l)+m)}{2 (2 N+1) \left(N^2+N+6\right)}$. The sensitivity for linear evolutions generated by $\hat{q}_{\mathbf{n}}=\mathbf{n}_{\mathrm{lin}}\cdot\hat{\mathbf{H}}$ with $\mathbf{n}_{\mathrm{lin}}=(n_1,n_2,0,0,0,0)$ is determined by the $2 \times 2$ principal submatrix
\begin{align}
\tilde{\mathbf{M}}[|N\rangle,\hat{\mathbf{H}}]=\begin{pmatrix} 
 4 N+2 & 0 \\ 0 & 4 N+2 \end{pmatrix}
\end{align}
in the top-left corner. This matrix is equivalent to the quantum Fisher matrix of Fock states, which describes their full sensitivity under displacements. Hence, we have shown that measurements of first- and third-order operators are optimal to capture the sensitivity of Fock states under displacements. We find $\chi_Q^{-2}[|N\rangle,\hat{q}_{\mathbf{n}},\hat{\mathbf{H}}]=4N + 2$ for all $\mathbf{n}_{\mathrm{lin}}$, as reported in Eq.~(\ref{eq:9}).


\begin{thebibliography}{}

\bibitem{Caves} C. M. Caves, Quantum-mechanical noise in an interferometer, \href{https://doi.org/10.1103/PhysRevD.23.1693}{Phys. Rev. D \textbf{23}, 1693 (1981)}.

\bibitem{Wineland} D. J. Wineland, J. J. Bollinger, W. M. Itano, F. L. Moore, and
D. J. Heinzen, Spin squeezing and reduced quantum noise in spectroscopy, \href{https://doi.org/10.1103/PhysRevA.46.R6797}{Phys. Rev. A \textbf{46}, R6797 (1992)}.

\bibitem{RMP} L. Pezz\`{e}, A. Smerzi, M. K. Oberthaler, R. Schmied, and P. Treutlein, 
Quantum metrology with nonclassical states of atomic ensembles, \href{https://doi.org/10.1103/RevModPhys.90.035005}{Rev. Mod. Phys. \textbf{90}, 035005 (2018)}.

\bibitem{TothJPA} G. T\'oth and I. Apellaniz,
Quantum metrology from a quantum information science perspective, 
\href{https://doi:10.1088/1751-8113/47/42/424006}{J. Phys. A \textbf{47}, 424006 (2014)}. 

\bibitem{Kimble} M. Xiao, L. A. Wu, and H. J. Kimble, 
Precision measurement beyond the shot-noise limit, \href{https://doi.org/10.1103/PhysRevLett.59.278}{Phys. Rev. Lett. \textbf{59}, 278 (1987)}.

\bibitem{GrangierPRL1987}
P. Grangier, R. E. Slusher, B. Yurke, and A. LaPorta, 
Squeezed-light-enhanced polarization interferometer
\href{https://doi.org/10.1103/PhysRevLett.59.2153}{Phys. Rev. Lett. \textbf{59}, 2153 (1987).}

\bibitem{KrusePRL2016}
K. Lange, J. Peise, B. L\"{u}cke, L. Pezz\`{e}, J. Arlt, W. Ertmer, C. Lisdat, L. Santos, A. Smerzi, and C. Klempt,
Improvement of an atomic clock using squeezed vacuum,
\href{https://doi.org/10.1103/PhysRevLett.117.143004}{Phys. Rev. Lett. \textbf{117}, 143004 (2016)}.

\bibitem{SchnabelNATCOMM2010}
R. Schnabel, N. Mavalvala, D. E. McClelland, and P. K. Lam,
Quantum metrology for gravitational wave astronomy, 
\href{https://doii:10.1038/ncomms1122}{Nat. Comm. \textbf{1}, 121 (2010)}.

\bibitem{AasiNATPHOT2013}
J. Aasi {\it et al.}, 
Enhanced sensitivity of the LIGO gravitational wave detector by using squeezed states of light, 
\href{https://doi.org/10.1038/nphoton.2013.177}{Nat. Photonics {\bf 7}, 613 (2013)}.

\bibitem{vanLoock}
P. van Loock and A. Furusawa, Detecting genuine multipartite continuous-variable entanglement, \href{https://doi.org/10.1103/PhysRevA.67.052315}{Phys. Rev. A \textbf{67}, 052315 (2003)}.

\bibitem{Adesso} G. Adesso and F. Illuminati, Entanglement in continuous variable systems: Recent advances and current perspectives, \href{https://doi.org/10.1088/1751-8113/40/28/S01}{J. Phys. A \textbf{40}, 7821 (2007)}.

\bibitem{GrossNATURE}
C. Gross, H. Strobel, E. Nicklas, T. Zibold, N. Bar-Gill, G. Kurizki, and M. K. Oberthaler, 
Atomic homodyne detection of continuous-variable entangled twin-atom states, 
\href{https://doi.org/10.1038/nature10654}{Nature \textbf{480}, 219 (2011)}.

\bibitem{Furusawa}
S. Yokoyama, R. Ukai, S. C. Armstrong, C. Sornphiphatphong, T. Kaji, S. Suzuki, J.-i. Yoshikawa, H. Yonezawa, N. C. Menicucci and A. Furusawa, Ultra-large-scale continuous-variable cluster states multiplexed in the time domain, \href{https://doi.org/10.1038/nphoton.2013.287}{Nat. Photon. \textbf{7}, 982 (2013)}.

\bibitem{Treps}
J. Roslund, R. Medeiros de Ara\'{u}jo, S. Jiang, C. Fabre and N. Treps, Wavelength-multiplexed quantum networks with ultrafast frequency combs, \href{https://doi.org/10.1038/nphoton.2013.340}{Nat. Photon. \textbf{8}, 109 (2014)}.

\bibitem{Quantum2017}
M. Gessner, L. Pezz\`e, and A. Smerzi,
Entanglement and squeezing in continuous-variable systems,
\href{https://doi.org/10.22331/q-2017-07-14-17}{Quantum {\bf 1}, 17 (2017)}.

\bibitem{ReidRMP2009} M. D. Reid, P. D. Drummond, W. P. Bowen, E. G. Cavalcanti, P. K. Lam, H. A. Bachor, U. L. Andersen, and G. Leuchs, \textit{Colloquium:} The Einstein-Podolsky-Rosen paradox: From concepts to applications, \href{https://doi.org/10.1103/RevModPhys.81.1727}{Rev. Mod. Phys. \textbf{81}, 1727 (2009)}.

\bibitem{EPRoptical} Z. Y. Ou, S. F. Pereira, H. J. Kimble, and K. C. Peng, Realization of the Einstein-Podolsky-Rosen paradox for continuous variables, \href{https://doi.org/10.1103/PhysRevLett.68.3663}{Phys. Rev. Lett. \textbf{68}, 3663 (1992)}.

\bibitem{EPRatomic} J. Peise, I. Kruse, K. Lange, B. L\"{u}cke, L. Pezz\`{e}, J. Arlt, W. Ertmer, K. Hammerer, L. Santos, A. Smerzi, and C. Klempt, Satisfying the Einstein-Podolsky-Rosen criterion with massive particles, \href{https://doi.org/10.1038/ncomms9984}{Nat. Commun. \textbf{6}, 8984 (2015)}.

\bibitem{FadelSCIENCE2018}
M. Fadel, T. Zibold, B. D\'ecamps, and P. Treutlein, 
Spatial entanglement patterns and Einstein-Podolsky-Rosen steering in Bose-Einstein condensates, 
\href{https://doi.org/10.1126/science.aao1850}{Science {\bf 360}, 409 (2018)}; 
P. Kunkel, M Pr\"ufer, H. Strobel, D. Linnemann, A. Fr\"olian, T. Gasenzer, M. G\"arttner, and M. K. Oberthaler, 
Spatially distributed multipartite entanglement enables EPR steering of atomic clouds, \href{https://doi.org/10.1126/science.aao2254}{Science {\bf 360}, 413 (2018)}.

\bibitem{LudlowRMP2015}
A. D. Ludlow, M. M. Boyd, J. Ye, E. Peik, and P. O. Schmidt,
Optical atomic clocks, 
\href{https://doi.org/10.1103/RevModPhys.87.637}{Rev. Mod. Phys. {\bf 87}, 637 (2015)}.

\bibitem{Mitchell} R. J. Sewell, M. Koschorreck, M. Napolitano, B. Dubost, N. Behbood, and M. W. Mitchell, Magnetic Sensitivity Beyond the Projection Noise Limit by Spin Squeezing, \href{https://doi.org/10.1103/PhysRevLett.109.253605}{Phys. Rev. Lett. \textbf{109}, 253605 (2012)}.

\bibitem{Pritchard}
A. D. Cronin, J. Schmiedmayer, and D. E. Pritchard, Optics and interferometry with atoms and molecules, \href{https://doi.org/10.1103/RevModPhys.81.1051}{Rev. Mod. Phys. \textbf{81}, 1051 (2009)}.


\bibitem{SMPRL01} A. S. S\o{}rensen and K. M\o{}lmer, Entanglement and Extreme Spin Squeezing, \href{http://dx.doi.org/10.1103/PhysRevLett.86.4431}{Phys. Rev. Lett. \textbf{86}, 4431 (2001)}.

\bibitem{Sorensen} A. S\o{}rensen, L. M. Duan, J. I. Cirac, and P. Zoller, Many-particle entanglement with Bose-Einstein condensates, \href{http://dx.doi.org/10.1038/35051038}{Nature \textbf{409}, 63 (2001)}.

\bibitem{Toth} G. T\'{o}th, C. Knapp, O. G\"{u}hne, and H. J. Briegel, Spin squeezing and entanglement, \href{https://doi.org/10.1103/PhysRevA.79.042334}{Phys. Rev. A \textbf{79}, 042334 (2009)}.

\bibitem{HyllusPRA2012b}
P. Hyllus, L. Pezz\`e, A. Smerzi, and Geza T\'oth,
Entanglement and extreme spin squeezing for a fluctuating number of indistinguishable particles, 
\href{https://doi.org/10.1103/PhysRevA.86.012337}{Phys. Rev. A {\bf 86}, 012337 (2012)}.

\bibitem{Tura} J. Tura, R. Augusiak, A. B. Sainz, T. V\'{e}rtesi, and M. Lewenstein, A. Ac\'{i}n, Detecting nonlocality in many-body quantum states, \href{https://doi.org/10.1126/science.1247715}{Science \textbf{344}, 1256 (2014)}.

\bibitem{Schmied} R. Schmied, J.-D. Bancal, B. Allard, M. Fadel, V. Scarani, P. Treutlein, and N. Sangouard, Bell correlations in a Bose-Einstein condensate, \href{https://doi.org/10.1126/science.aad8665}{Science \textbf{352}, 441 (2016)}.

\bibitem{EngelsenPRL2017}
N. J. Engelsen, R. Krishnakumar, O. Hosten, and M. A. Kasevich,
Bell correlations in spin-squeezed states of 500 000 atoms, \href{https://doi.org/10.1103/PhysRevLett.118.140401}{Phys. Rev. Lett. {\bf 118}, 140401 (2017)}.

\bibitem{KitagawaUeda} M. Kitagawa and M. Ueda, Squeezed spin states, \href{https://doi.org/10.1103/PhysRevA.47.5138}{Phys. Rev. A \textbf{47}, 5138 (1993)}.


\bibitem{Riedel} C. Gross, T. Zibold, E. Nicklas, J. Est\`{e}ve, and M. K. Oberthaler, Nonlinear atom interferometer surpasses classical precision limit, \href{https://doi.org/10.1038/nature08919}{Nature \textbf{464}, 1165 (2010)}; M. F. Riedel, P. B\"{o}hi, Y. Li, T. W. H\"{a}nsch, A. Sinatra, and P. Treutlein, Atom-chip-based generation of entanglement for quantum metrology, \href{https://doi.org/10.1038/nature08988}{Nature \textbf{464}, 1170 (2010)}.

\bibitem{Sinatra} A. Sinatra, E. Witkowska, J.-C. Dornstetter, Y. Li, and Y. Castin, Limit of Spin Squeezing in Finite-Temperature Bose-Einstein Condensates, \href{https://doi.org/10.1103/PhysRevLett.107.060404}{Phys. Rev. Lett. \textbf{107}, 060404 (2011)}.

\bibitem{Bohnet} J. G. Bohnet, B. C. Sawyer, J. W. Britton, M. L. Wall, A. M. Rey, M. Foss-Feig, J. J. Bollinger, Quantum spin dynamics and entanglement generation with hundreds of trapped ions, \href{https://doi.org/10.1126/science.aad9958 }{Science \textbf{352}, 1297 (2016)}.

\bibitem{LerouxPRL2010}
I.D. Leroux, M. H. Schleier-Smith, and V. Vuleti\`c, 
Implementation of cavity squeezing of a collective atomic spin, 
\href{https://doi.org/10.1103/PhysRevLett.104.073602}{Phys. Rev. Lett. {\bf 104}, 073602 (2010)}.

\bibitem{Hosten} O. Hosten, N. J. Engelsen, R. Krishnakumar and M. A. Kasevich, Measurement noise 100 times lower than the quantum-projection limit using entangled atoms, \href{https://doi.org/10.1038/nature16176}{Nature \textbf{529}, 505 (2016)}; 
K. C. Cox, G. P. Greve, J. M. Weiner, and J. K. Thompson, 
Deterministic squeezed states with collective measurements and feedback, \href{https://doi.org/10.1103/PhysRevLett.116.093602}{Phys. Rev. Lett. {\bf 116}, 093602 (2016)}.

\bibitem{LocalSqueezing} M. Gessner, L. Pezz\`{e}, and A. Smerzi, Resolution-enhanced entanglement detection, \href{https://doi.org/10.1103/PhysRevA.95.032326}{Phys. Rev. A \textbf{95}, 032326 (2017)}.

\bibitem{Ma} J. Ma, X. Wang, C. Sun, and F. Nori, 
Quantum spin squeezing, \href{https://doi.org/10.1016/j.physrep.2011.08.003}{Phys. Rep. \textbf{509}, 89 (2011)}.

\bibitem{BraunsteinVanLoock} S. L. Braunstein and P. van Loock, Quantum information with continuous variables, \href{https://doi.org/10.1103/RevModPhys.77.513}{Rev. Mod. Phys. \textbf{77}, 513 (2005)}.

\bibitem{Ferraro} A. Ferraro, S. Olivares, and M. G. A. Paris, \href{http://arxiv.org/abs/quant-ph/0503237}{Gaussian states in continuous variable quantum information}, (Bibliopolis, Napoli, 2005).

\bibitem{Wang} X. Wang, T. Hiroshima, A. Tomita, and M. Hayashi, Quantum information with Gaussian states, \href{https://doi.org/10.1016/j.physrep.2007.04.005}{Phys. Rep. \textbf{448}, 1 (2007)}.

\bibitem{Weedbrook} C. Weedbrook, S. Pirandola, R. Garc\'{i}a-Patr\'o{o}n, N. J. Cerf, T. C. Ralph, J. H. Shapiro, and S. Lloyd, Gaussian quantum information, \href{https://doi.org/10.1103/RevModPhys.84.621}{Rev. Mod. Phys. \textbf{84}, 621 (2012)}.

\bibitem{Lang} M. D. Lang and C. M. Caves, Optimal Quantum-Enhanced Interferometry Using a Laser Power Source, \href{https://doi.org/10.1103/PhysRevLett.111.173601}{Phys. Rev. Lett. \textbf{111}, 173601 (2013)}.

\bibitem{Zurek}
W. H. Zurek, Sub-Planck structure in phase space and its relevance for quantum decoherence, \href{https://doi.org/10.1038/35089017}{Nature \textbf{412}, 712 (2001)}.

\bibitem{Ourjoumtsev} A. Ourjoumtsev, H. Jeong, R. Tualle-Brouri, and P. Grangier, Generation of optical `Schr\"{o}dinger cats' from photon number states, \href{https://doi.org/10.1038/nature06054}{Nature \textbf{448}, 784 (2007)}.

\bibitem{DelegliseNATURE2008} S. Del\'{e}glise, I. Dotsenko, C. Sayrin, J. Bernu, M. Brune, J.-M. Raimond, and S. Haroche, Reconstruction of non-classical cavity field states with snapshots of their decoherence, \href{https://doi.org/10.1038/nature07288}{Nature \textbf{455}, 510 (2008)}.

\bibitem{Wolf} F. Wolf, C. Shi, J. C. Heip, M. Gessner, L. Pezz\`{e}, A. Smerzi, M. Schulte, K. Hammerer, and P. O. Schmidt, Motional Fock states for quantum-enhanced amplitude and phase measurements with trapped ions, \href{https://arxiv.org/abs/1807.01875}{arXiv:1807.01875}.

\bibitem{PS09} L. Pezz\'{e} and A. Smerzi, Entanglement, Nonlinear Dynamics, and the Heisenberg Limit, \href{https://link.aps.org/doi/10.1103/PhysRevLett.102.100401}{Phys. Rev. Lett. \textbf{102}, 100401 (2009)}.

\bibitem{GiovannettiNATPHOT2011}
V. Giovannetti, S. Lloyd and L. Maccone, 
Advances in quantum metrology, 
\href{https://doi.org/10.1038/nphoton.2011.35}{Nat. Phot. {\bf 5}, 222 (2011)}.

\bibitem{Varenna}
L. Pezz\`e and A. Smerzi, 
Quantum theory of phase estimation, \textit{in Atom Interferometry, Proceedings of the International School of Physics ``Enrico Fermi'', Course 188, Varenna}, edited by G. M. Tino and M. A. Kasevich (IOS Press, Amsterdam) p. 691 (2014).

\bibitem{Bartlett} S. D. Bartlett, B. C. Sanders, S. L. Braunstein, and K. Nemoto, Efficient Classical Simulation of Continuous Variable Quantum Information Processes, \href{https://doi.org/10.1103/PhysRevLett.88.097904}{Phys. Rev. Lett. \textbf{88}, 097904 (2002)}.

\bibitem{Eisert}
J. Eisert, S. Scheel, and M. B. Plenio, Distilling Gaussian States with Gaussian Operations is Impossible, \href{https://doi.org/10.1103/PhysRevLett.89.137903}{Phys. Rev. Lett. \textbf{89}, 137903 (2002)};  J. Fiur\'{a}\v{s}ek, Gaussian transformations and distillation
of entangled Gaussian states, \href{https://doi.org/10.1103/PhysRevLett.89.137904}{Phys. Rev. Lett. \textbf{89}, 137904 (2002)}.

\bibitem{Giedke} G. Giedke and J. I. Cirac, Characterization of Gaussian operations and distillation of Gaussian states, \href{https://doi.org/10.1103/PhysRevA.66.032316}{Phys. Rev. A \textbf{66}, 032316 (2002)}.

\bibitem{Andersen} U. L. Andersen, J. S. Neergaard-Nielsen, P. van Loock, and A. Furusawa, Hybrid discrete- and continuous-variable quantum information, \href{https://doi.org/10.1038/nphys3410}{Nat. Phys. \textbf{11}, 713 (2015)}.

\bibitem{Strobel} H. Strobel, W. Muessel, D. Linnemann, T. Zibold, D. B. Hume, L. Pezz\`{e}, A. Smerzi, M. K. Oberthaler, Fisher information and entanglement of non-Gaussian spin states, \href{https://doi.org/10.1126/science.1250147}{Science \textbf{345}, 424 (2014)}.

\bibitem{Sciarrino} A. Crespi, R. Osellame, R. Ramponi, D. J. Brod, E. F. Galv\~{a}o, N. Spagnolo, C. Vitelli, E. Maiorino, P. Mataloni, and F. Sciarrino, Integrated multimode interferometers with arbitrary designs for photonic boson sampling, \href{https://doi.org/10.1038/nphoton.2013.112}{Nat. Photon. \textbf{7}, 545 (2013)}.

\bibitem{Grangier} J. Wenger, R. Tualle-Brouri, and P. Grangier, Non-Gaussian Statistics from Individual Pulses of Squeezed Light, \href{https://doi.org/10.1103/PhysRevLett.92.153601}{Phys. Rev. Lett. \textbf{92}, 153601 (2004)}.

\bibitem{Bellini} V. Parigi, A. Zavatta, M. Kim, and M. Bellini, Probing Quantum Commutation Rules by Addition and Subtraction of Single Photons to/from a Light Field, \href{https://doi.org/10.1126/science.1146204}{Science \textbf{317}, 1890 (2007)}.

\bibitem{Walschaers} M. Walschaers, C. Fabre, V. Parigi, and N. Treps, Entanglement and Wigner Function Negativity of Multimode Non-Gaussian States, \href{https://doi.org/10.1103/PhysRevLett.119.183601}{Phys. Rev. Lett. \textbf{119}, 183601 (2017)}.

\bibitem{HaasSCIENCE2014} F. Haas, J. Volz, R. Gehr, J. Reichel and J. Est\'{e}ve, Entangled States of More Than 40 Atoms in an Optical Fiber Cavity, \href{10.1126/science.1248905}{Science \textbf{344}, 180 (2014)}.

\bibitem{KimPRA1998}
T. Kim, O. Pfister, M. J. Holland, J. Noh, and J. L. Hall, 
Influence of decorrelation on Heisenberg-limited interferometry with quantum correlated photons, 
\href{https://doi.org/10.1103/PhysRevA.57.4004}{Phys. Rev. A {\bf 57}, 4004 (1998)}.

\bibitem{TF} B. L\"{u}cke, M. Scherer, J. Kruse, L. Pezz\`{e}, F. Deuretzbacher, P. Hyllus, O. Topic, J. Peise, W. Ertmer, J. Arlt, L. Santos,
A. Smerzi, and C. Klempt, Twin Matter Waves for Interferometry Beyond the Classical Limit, \href{http://dx.doi.org/10.1126/science.1208798}{Science \textbf{334}, 773 (2011)}.

\bibitem{BollingerPRA1996}
J. J. Bollinger, W. M. Itano, D. J. Wineland, and D. J. Heinzen, 
Optimal frequency measurements with maximally correlated states,
\href{https://doi.org/10.1103/PhysRevA.54.R4649}{Phys. Rev. A {\bf 54}, R4649 (1996)}.

\bibitem{LeibfriedNATURE2005}
D. Leibfried, E. Knill, S. Seidelin, J. Britton, R. B. Blakestad, J. Chiaverini, D. B. Hume, W. M. Itano, J. D. Jost, C. Langer, R. Ozeri, R. Reichle, and D. J. Wineland, 
Creation of a six-atom ``Schr\"{o}dinger cat'' state, 
\href{https://doi.org/10.1038/nature04251}{Nature {\bf 438}, 639 (2005)}.

\bibitem{MonzPRL2011}
T. Monz, P. Schindler, J. T. Barreiro, M. Chwalla, D. Nigg, W. A. Coish, M. Harlander, W. H\"ansel, M. Hennrich, and R. Blatt, 
14-qubit entanglement: Creation and coherence, 
\href{https://doi.org/10.1103/PhysRevLett.106.130506}{Phys. Rev. Lett. {\bf 106}, 130506 (2011)}.

\bibitem{BraunsteinPRL1994}
S. L. Braunstein and C. M. Caves,
Statistical distance and the geometry of quantum states
\href{https://doi.org/10.1103/PhysRevLett.72.3439}{Phys. Rev. Lett. {\bf 72}, 3439 (1994)}.

\bibitem{Kholevo}
A. S. Kholevo, A Generalization of the Rao-Cramer Inequality, \href{https://doi.org/10.1137/1118039}{Theory Probab. Appl. \textbf{18}, 359 (1974)}.

\bibitem{Frowis2015} F. Fr\"{o}wis, R. Schmied, and N. Gisin, Tighter quantum uncertainty relations following from a general probabilistic bound, \href{https://doi.org/10.1103/PhysRevA.92.012102}{Phys. Rev. A \textbf{92}, 012102 (2015)}.


\bibitem{HelstromBOOK}
C.M. Helstrom,
{\it Quantum Detection and Estimation Theory} (Academic Press, 1976).

\bibitem{note1} In general, $F_Q[\hat{\rho}(\theta)]=\mathrm{Tr}\{\hat{\rho}(\theta)\hat{L}^2\}$, where the symmetric logarithmic derivative operator $\hat{L}$ may depend on $\theta$ and is defined by $d\hat{\rho}(\theta)/d\theta=[\hat{L}\hat{\rho}(\theta)+\hat{\rho}(\theta)\hat{L}]/2$. For unitary evolutions $F_Q[\hat{\rho}(\theta)]=F_Q[\hat{\rho},\hat{H}]$ depends only on $\hat{\rho}$ and $\hat{H}$.

\bibitem{note2} For unitary evolutions, we consider the case $\theta=0$ for ease of notation and replace the argument $\hat{\rho}(\theta)$ by the initial state $\hat{\rho}$ and the Hamiltonian $\hat{H}$.

\bibitem{Supp} See the Supplementary Material, which contains Refs.~\cite{Kholevo,Frowis2015,BraunsteinPRL1994}, for a proof of the convexity property, the saturation of the quantum Fisher information by an optimal observable and details on the nonlinear squeezing coefficients for collective spins and Fock states.




\bibitem{MandelWolf} L. Mandel and E. Wolf, \textit{Optical Coherence and Quantum Optics}, (Cambridge University Press, Cambridge, UK, 1995).

\bibitem{Luis} \'{A}. Rivas and A. Luis, Precision Quantum Metrology and Nonclassicality in Linear and Nonlinear Detection Schemes, \href{https://doi.org/10.1103/PhysRevLett.105.010403}{Phys. Rev. Lett. \textbf{105}, 010403 (2010)}.

\bibitem{HyllusPRA2012}
P. Hyllus {\it et al.}, 
Fisher information and multiparticle entanglement, 
\href{https://doi.org/10.1103/PhysRevA.85.022321}{Phys. Rev. A {\bf 85}, 022321 (2012)}; 
G. T\'oth, 
Multipartite entanglement and high-precision metrology, 
\href{https://doi.org/10.1103/PhysRevA.85.022322}{Phys. Rev. A {\bf 85}, 022322 (2012)}.

\bibitem{note4} In fact a sharper bound for $N$-qubit states with up to $k$ entangled qubits \cite{HyllusPRA2012} is given as $\xi^{-2}_{\rm opt}[\hat{\rho},\hat{H},\hat{\mathbf{H}}]\leq sk^2+r^2$ where $s=\lfloor N/k\rfloor$ and $r=N-sk$.

\bibitem{note5}
For instance $\{\hat{J}_x,\hat{J}_y\}$ can be accessed by measuring $(\hat{J}_x + \hat{J}_y)^2/2$ and $(\hat{J}_x - \hat{J}_y)^2/2$.


\bibitem{MolmerPRL1999} K. M\o{}lmer and A. S\o{}rensen, Multiparticle Entanglement of Hot Trapped Ions, \href{https://doi.org/10.1103/PhysRevLett.82.1835}{Phys. Rev. Lett. \textbf{82}, 1835 (1999)}.

\bibitem{ChalopinArXiv}
T. Chalopin, C. Bouazza, A. Evrard, V. Makhalov, D. Dreon, J. Dalibard, L. A. Sidorenkov, and S. Nascimbene, 
Quantum-enhanced sensing using non-classical spin states of a highly magnetic atom, 
\href{https://doi.org/10.1038/s41467-018-07433-1}{Nat. Comm. {\bf 9}, 4955 (2018)}.

\bibitem{footnoteParity} The spin-parity-squeezing coefficient can be further enhanced by optimizing the phase shift $\theta$ for states at $\tau<\pi/2$.

\bibitem{Muessel} W. Muessel, H. Strobel, D. Linnemann, T. Zibold, B. Juli\'{a}-D\'{i}az, and M. K. Oberthaler, Twist-and-turn spin squeezing in Bose-Einstein condensates, \href{https://doi.org/10.1103/PhysRevA.92.023603}{Phys. Rev. A \textbf{92}, 023603 (2015)}.

\bibitem{Sorelli} G. Sorelli, M. Gessner, A. Smerzi and L. Pezz\`{e}, Fast and optimal generation of entanglement in bosonic Josephson junctions, \href{https://doi.org/10.1103/PhysRevA.99.022329}{Phys. Rev. A \textbf{99}, 022329 (2019)}.


\bibitem{note6} Reduced quantum fluctuations of higher-order moments of the electromagnetic field have been pointed out in 
C. K. Hong and L. Mandel, Higher-Order Squeezing of a Quantum Field, \href{https://doi.org/10.1103/PhysRevLett.54.323}{Phys. Rev. Lett. \textbf{54}, 323 (1985)}. The nonlinear squeezing coefficients, defined in Eq.~(\ref{eq:xiQ}), optimize such higher-order moments and quantify their metrological sensitivity.

\bibitem{Leibfried} D. Leibfried, R. Blatt, C. Monroe, and D. Wineland, Quantum dynamics of single trapped ions, \href{https://doi.org/10.1103/RevModPhys.75.281}{Rev. Mod. Phys. \textbf{75}, 281 (2003)}.

\bibitem{Paulisch} V. Paulisch, M. Perarnau-Llobet, A. Gonz\'{a}lez-Tudela, J. I. Cirac, Quantum metrology with one-dimensional superradiant photonic states, \href{https://arxiv.org/abs/1805.00712}{arXiv:1805.00712}.







\newpage
\begin{center}
{\large \bfseries Supplementary Material}
\end{center}

\setcounter{equation}{0} \setcounter{figure}{0} \renewcommand\thefigure{S%
\arabic{figure}} \renewcommand\theequation{S\arabic{equation}}
\end{thebibliography}
\end{document}